\documentclass[5p,times, review]{elsarticle}

\usepackage[T1]{fontenc} % Use modern font encodings

\usepackage{xcolor}
\usepackage{amsmath}
\usepackage{graphicx}

\begin{document}

\begin{frontmatter}
  \title{Single-layer of Bi$_{1-x}$Sb$_x$ grown on Ag(111).}

  \author[1]{Javier D. Fuhr}
  \ead{fuhr@cab.cnea.gov.ar}

  \author[2]{J. Esteban Gayone}

  \author[2]{Hugo Ascolani}
  \ead{ascolani@cab.cnea.gov.ar}

  \address[1]{Centro Atómico Bariloche, CNEA y CONICET, Av. Bustillo 9500, (8400) S.C. de Bariloche (RN), Argentina.}

  \address[2]{Instituto de Nanociencia y Nanotecnología (CNEA - CONICET), Nodo Bariloche, Av. Bustillo 9500, (8400) S.C. de Bariloche (RN), Argentina.}

  \date{\today}

  \begin{abstract}

    In this work, we report the growth of a single mixed Bi$_{(1-x)}$ Sb$_x$ layer, with diverse stoichiometries, on a Ag(111) substrate. The atomic geometry has been thoroughly investigated by low energy electron diffraction, scanning tunneling microscopy, and X-ray photoelectron spectroscopy experiments, as well as calculations based on density functional theory (DFT).
    We first determined that both pure systems (Bi/Ag(111) and Sb/Ag(111)) show similar behaviors: they form surface alloys with  $(\sqrt{3} \times \sqrt{3})R30^{\circ}$ periodicity for coverages lower than 1/3~ML, and undergo a dealloying transition for higher coverages up to 2/3~ML.
    We then established a simple preparation procedure to obtain a mixed Bi-Sb overlayer on Ag(111): it is essential to start with a surface completely covered by either of the two pure surface alloys and then deposit the other element on it.
    The energetics derived from DFT calculations provide insight into the system's preference towards the formation of this phase, and also predict a pathway to the formation of Bi-rich non-alloyed phases.
    The obtained mixed Bi-Sb phase has a lateral atomic arrangement very similar to the one in the non-alloyed phase observed for Sb on Ag(111), with Sb and Bi atoms distributed disorderly, and presents a significant vertical corrugation, promising considerable Rashba effects.
  \end{abstract}
\end{frontmatter}

\section{\label{intro}Introduction}

Surfaces with a strong spin-textured electronic structure are of great interest as a platform for building spintronic devices.\cite{Rojas2013,Manchon2015} In systems without inversion symmetry, the spin-orbit (SO) interaction causes the breaking of the spin degeneracy of the electronic states, giving rise to two bands, each one with opposite spin orientation. The SO splitting is proportional to the gradient of the electrostatic potential undergone by the electrons. The requisites for a considerable effect could be synthesized as follows: (i) a long-range ordered surface to maintain 2D electronic states, (ii) comparably light atoms surrounding heavy atoms, and (iii) vertical corrugation.

A notable example is the $(\sqrt{3} \times \sqrt{3})R30^{\circ}$ Bi/Ag(111) surface alloy (BiAg$_2$ surface alloy, hereafter), made by depositing $1/3$ monolayer (ML) of Bi on the Ag(111) surface, which has received much attention due to its giant Rashba effect.\cite{Ast2007,Pacile2006} The origin of the observed giant spin splitting in this surface alloy has been attributed to a near-optimal surface corrugation and a large atomic spin-orbit coupling (SOC) in Bi.\cite{Bian2013}  The properties of the BiAg$_2$ surface alloy triggered the investigation of the $(\sqrt{3} \times \sqrt{3})R30^{\circ}$ Sb/Ag(111) surface alloy (SbAg$_2$ surface alloy, hereafter)\cite{Moreschini2009} one and also of mixed surface alloys including the  Bi$_{(1-x)}$Sb$_x$Ag$_2$/Ag(111) one.\cite{Gierz2011,Meier2009}

Bismuth and antimony crystallize with rhombohedral symmetry in a structure which is typical for the group V semimetals (structure A7, space group R$\bar{3}m$). The band structure of bulk bismuth is drastically modified by the substitution of Bi atoms by Sb atoms, to such an extent that single crystals of bismuth antimonide (Bi$_{(1-x)}$Sb$_x$), with $x$ between $7\%$ and $30\%$, present a topological insulator phase.\cite{Hsieh2008,Lemaitre2022}

The electronic properties of two-dimensional BiSb monolayers have been theoretically investigated, and the results predict a strong Rashba spin splitting of the electronic states. \citep{BiSb2010,BiSbAlN2017} However, no experimental studies on this class of binary systems have been reported. Therefore, motivated by the potential for exotic electronic properties that a monolayer of Bi$_{(1-x)}$ Sb$_x$ could have, we addressed the problem of synthesizing it in our laboratory. We describe below a preparation procedure to grow a single overlayer of Bi$_{(1-x)}$ Sb$_x$ on the Ag(111) substrate, with varying stoichiometry. We characterize the geometrical structure of the obtained Bi-Sb binary monolayer using low-energy electron diffraction (LEED), scanning tunneling microscopy (STM) and X-ray photoelectron spectroscopy (XPS), combined with calculations based on density functional theory (DFT).

\section{Experimental and calculation details}

Two independent ultra-high vacuum  systems for surface analysis were used in this investigation: one for the STM/LEED experiments and another for the XPS/UPS experiments.

The STM experiments were carried out in an ultra-high vacuum system equipped with a variable-temperature STM from Omicron (model AFM/STM VT 25 DRH), a LEED optics, and a homemade ion gun for Ar$^+$ sputtering. All the LEED images were obtained with an acceleration voltage of 5 KeV. The base pressure of the chamber was in the low $10^{-10}$ mBar range. All the reported STM images were taken at room temperature (RT) with tungsten (W) tips. Negative sample bias voltages correspond to occupied-states images.  The thermal drift was compensated during the measurements by applying the facility provided by the MATRIX software used to control the STM.

The XPS/UPS experiments, on the other hand, were conducted in a surface analysis system from SPECS with a base pressure in the main chamber in the low $10^{-10}$ mBar range. In this case, the two evaporation sources were mounted in the auxiliary chamber (base pressure of $5\times 10^{-9}$ mBar). Photoemission-spectroscopy data were acquired using a Phoibos 150 (SPECS GmbH) electron energy analyzer, utilizing photons from a monochromatic Al$_{K\alpha}$ source (1486.7 eV) or a He discharge lamp (21.22 eV). Photoemission spectra of the Sb-3d, Ag-3d, and Bi-4f core levels were systematically measured with a pass energy of 10 eV, resulting in an estimated energy resolution (analyzer plus X-rays) of 0.5 eV, with electrons collected along the normal to the surface. The Binding Energy (BE) scale of the XPS spectra was calibrated by setting the BE position of the bulk Ag-3d$_{5/2}$ peak to 368.26 eV. The UPS data were measured with a pass energy of 4 eV. The  Sb-3d$_{5/2}$, Bi-4f$_{7/2}$ and Ag-3d spectra were fitted using  Doniach-Sunjic (D-S) line shapes combined with Shirley-type backgrounds.  For both experimental systems, specific experiments (using an auxiliary sample) were performed to calibrate the sample temperature relative to the temperature measured by the sensors.

The Ag(111) substrate (Mateck, orientation accuracy $\pm 0.1$) was prepared by repeated cycles of Ar$^+$ at 1.5~KeV bombardment and subsequently annealed at 750~K. The Sb and Bi atoms  were evaporated from two independent  home-made Knudsen cells composed of resitively-heated boron-nitride crucibles. In the STM chamber, the evaporation rates of both sources were calibrated by  optimizing the  LEED  patterns of the respective surface alloys. Alternatively, in the case of the SPECS system, the evaporation rates were calibrated by optimizing the corresponding valence-band peaks obtained from angle-resolved UPS experiments. Specifically, we used evaporation rates of 0.05 ML/minute approximately.  The calibrations were checked systematically. In this study, the deposition of 1 ML is defined as the density of Ag atoms in a Ag(111)
plane (13.87 atoms per nm${^2}$).

The surface alloys were prepared according to the following procedure: both Bi and Sb atoms were evaporated on the Ag(111) substrate maintained at $90^\circ$C, and then the resulting samples were annealed at $150^\circ$C for 15 minutes.  On the other hand,  the mixed samples (Bi+Sb)/Ag(111) were obtained as follows: once  the chosen surface alloy was prepared,  the other element was deposited on it with the sample kept at $90^\circ$C.

The density functional theory (DFT) calculations were performed using the Quantum ESPRESSO package,\cite{Q-E} which is a plane-wave implementation with pseudopotentials. To take into account van der Waals interactions, which can be important for a 2D system on a metallic surface, we employed the vdW-DF2-B86R functional of Hamada\cite{Hamada2014}, which is a revised second version of the non-local vdW-DF functional.\cite{vdw-df2} We used a wave function/density cutoff of 32/256 Ry, and Brillouin integrations were conducted using a uniform $\Gamma$-centered $k$-point mesh of $30\times 30\times 30$ for bulk Ag. With these parameters, we obtained an optimized lattice parameter for Ag bulk of 4.096~\AA, which we used in the subsequent of the calculations. The convergence threshold for electronic self-consistency was set to 10$^{-7}$ a.u., while the convergence threshold on total energy for ionic minimization was set to 10$^{-5}$ a.u.

For the surface calculations, we used the slab method with eight pure Ag layers and the top layer which, depending on the case, corresponds to the surface alloy or the non-alloyed overlayer. We used a vacuum layer larger than 15~\AA\ between replicas along the $z$ direction to minimize interactions. In all the calculations, we fixed the two lower layers while all other atoms are allowed to relax. Being consistent with the bulk calculations, for the $(3 \times \sqrt{3})$ surface unit cell, we used for Brillouin integrations a uniform $\Gamma$-centered $k$-point mesh of $10\times 17\times 1$.

\section{\label{Resu}Results}

\subsection{Behaviour of the pure systems: Bi/Ag(111) and Sb/Ag(111)}

The growth of Bi on the Ag(111) substrate has been widely studied and is now well understood.\cite{ApSS2009,Pussi2011} For Bi coverages lower than 1/3~ML, the Bi/Ag(111) system forms the prototypical surface alloy BiAg$_2$/Ag(111) with $(\sqrt{3} \times \sqrt{3})R30^{\circ}$ periodicity. Increasing the Bi coverage beyond 1/3~ML leads to a dealloying transition and the formation of a well ordered overlayer of Bi atoms with an incommensurate $p \times \sqrt{3}$ structure, with $p > 3$.\cite{Pussi2011} Fig.~\ref{LEEDpuros} (a) shows a typical $(p \times \sqrt{3})$ LEED pattern of this structure. The dealloying process involves the substitution of the Ag atoms in the surface alloy by the arrived Bi atoms, resulting in $(p \times \sqrt{3})$ domains supported on the same Ag(111) layer as the initial BiAg$_2$ surface alloy.\cite{Pussi2011} Remarkably, this dealloying transition causes a position shift of the Bi-4f core level by 0.2 eV towards higher binding energies (BEs) and can, therefore, be detected by means of the XPS technique.\cite{Pussi2011}

On the other hand, similar to the Bi/Ag(111) system, the Sb/Ag(111) system also forms an SbAg$_2$ surface-alloy phase for Sb coverages lower than 1/3ML. However, its behavior for Sb coverages beyond 1/3ML is still controversial.
%On the other hand, while initially the deposition of Sb on Ag(111) for Sb coverages lower than 1/3~ML also forms a $(\sqrt{3} \times \sqrt{3})R30^{\circ}$ surface alloy,\cite{Moreschini2009} the nature of the phases formed for Sb coverages beyond 1/3~ML are still controversial.
While the formation of graphene-like antimonene (honeycomb structure) on Ag(111) has been claimed,{\cite{Shao2018, Mao2018} a very recent study based on Raman-spectroscopy and STM experiments has instead claimed the growth of bilayers of antimonene with phosphorene-like puckered structure, supported on the SbAg$_2$ surface alloy.\cite{Zhang2022} We therefore decided to first revise the growth of Sb on the Ag(111) surface in the range of Sb coverages between $1/3$ and $2/3$~ML.

  Figure~\ref{LEEDpuros}(b) shows a LEED pattern obtained from a Ag(111) surface  with 1/2~ML of Sb. The sample was prepared by maintaining the Ag(111) surface at $90^\circ$C during evaporation and then holding the sample at that temperature for an additional 1~hour, a procedure similar to the one reported in reference [\citenum{Zhang2022}]. The obtained LEED pattern is composed of well-defined spots and is compatible with both hexagonal $(2\sqrt{3} \times 2\sqrt{3})R30^{\circ}$ and rectangular $(3 \times \sqrt{3})$ periodicities. This ambiguity is resolved by the high-resolution STM image of Figs~\ref{LEEDpuros}(c), which shows a rectangular $(3 \times \sqrt{3})$ structure with four spots per unit cell. Contrary to the stable $(p \times \sqrt{3})$ phase of Bi/Ag(111), the $(3 \times \sqrt{3})$  structure of Sb/Ag(111) is difficult to obtain because the system tends to form disordered amorphous structures. See Section S1 of Supplemental Material for additional details.

\begin{figure}
\includegraphics[width=1.0\columnwidth] {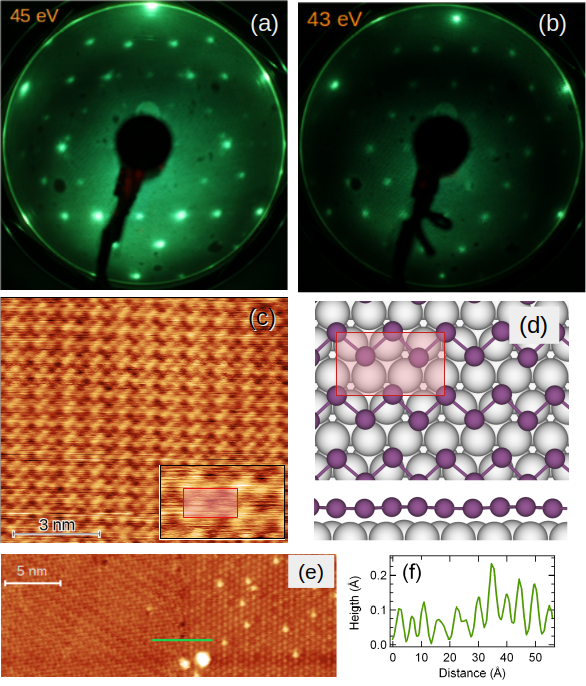}
 \caption{\label{LEEDpuros} (a):  LEED pattern taken at 45 eV of the non-commensurate  $(p \times \sqrt{3})$ structure of  Bi/Ag(111). (b):  LEED pattern measured at 43 eV of the commensurate  $(3 \times \sqrt{3})$ structure formed by 1/2 ML of Sb deposited on Ag(111). (c): STM image of the prepared Sb/Ag(111) surface.  Size: $(10 \times 10) $\AA$^2 $; Tunneling Conditions: (0.5V/1.5nA).  The inset shows a zoom of this image where  the $(3 \times \sqrt{3})$ unit cell containing four  protrusions is indicated.  (d): Top and side views of the unalloyed $(3 \times \sqrt{3})$ structure, with 4 Sb atoms inside the unit cell, obtained after a total-energy DFT relaxation (see Section S7 of Supplemental Material for details). Purple (grey) circles represent Sb (Ag) atoms. (e) STM image with islands of the $(3 \times \sqrt{3})$ Sb/Ag(111) and the SbAg$_2$ surface alloy. T. Cond.:  (-0.4V/1.4nA). (f)Height profile along to the green line indicated in (e). The average height difference between the two phases is 0.06\AA. }
 \end{figure}

 Although these experimental results agree with those reported in [\citenum{Zhang2022}], we obtained clear evidence that the $(3 \times \sqrt{3})$ structure does not grow on top of the SbAg$_2$ surface alloy as proposed in that work. In fact, the STM images where islands of the $(3 \times \sqrt{3})$ structure appeared adjacent to the SbAg$_2$ surface alloy (see Figs~\ref{LEEDpuros}(e) and \ref {LEEDpuros}(f)), demonstrate that this phase consists of a monolayer with a rectangular structure resting directly on the Ag(111) substrate. Based on these observations, we propose that the obtained $(3 \times \sqrt{3})$ phase is analogous to the one formed in the Sb/Au(111) system \cite{Cantero2021}, i.e. a monolayer that forms zig-zag chains along the $<\bar{1}10>$ axes of the Ag(111) substrate, as illustrated in Fig.~\ref{LEEDpuros}(d). Interestingly, the $(p \times \sqrt{3})$ structure of the Bi/Ag(111) system is similar to this one with the exception that, due to the larger size of the Bi atoms, the zig-zag chains they are not commensurate with the Ag(111) substrate.

 % To investigate  the thermal stability of the $(3 \times \sqrt{3})R30^{\circ}$ phase of the Sb/Ag(111) system, we monitored the evolution with temperature of the LEED pattern, by annealing the sample at a given temperature and then measuring the corresponding pattern.
 To investigate the thermal stability of the $(3 \times \sqrt{3})R30^{\circ}$ phase of the Sb/Ag(111) system, we monitored the evolution of the LEED pattern with sample temperature, i.e., the sample was annealed at a given temperature, and then the corresponding pattern was measured.
 We found a transition from the initial $(3 \times \sqrt{3})$ pattern to a $(\sqrt{3} \times \sqrt{3})R30^{\circ}$ one, accompanied by the complete disappearance of the $(3 \times \sqrt{3})$ spots at $(240\pm 20)^\circ$C.
This transition strongly suggests the activation of a pathway for the Sb atoms to migrate towards the Ag bulk, allowing a reduction of the Sb atoms on the surface, the disassembly of the rectangular structure and the reappearance of the surface alloy.

In summary, we found similar behaviors of Bi and Sb when deposited on the Ag(111) surface. Below 1/3~ML, both systems form stable surface alloys, and when coverage exceeds 1/3~ML they undergo a dealloying transition generating pure monolayers supported on the Ag(111) substrate with similar rectangular structures. Therefore, it is reasonable to expect that the coadsorption of Bi and Sb on Ag(111) could also generate rectangular structures of mixed Bi-Sb composition.

\subsection{Formation of mixed Bi-Sb structures}

In order to obtain mixed Bi-Sb phases we performed several experiments in which we started with the $(\sqrt{3} \times \sqrt{3})R30^{\circ}$ surface alloy of one of the elements and then deposited the other one.

Firstly, we started with 1/6~ML Sb/Ag(111) and deposited Bi. Up to 1/3~ML of total Sb+Bi coverage, we continued observing a $(\sqrt{3} \times \sqrt{3})R30^{\circ}$ phase, which we identify as a mixed Bi$_{(1-x)}$Sb$_x$Ag$_2$ surface alloy.\cite{Gierz2011} When the total deposition of Bi+Sb exceeds 1/3~ML, we see a change in the LEED pattern indicating the formation of the non-alloyed $(p \times \sqrt{3})$ phase of the pure Bi/Ag(111). Moreover, STM images obtained for the particular case of 1/3~ML of Bi deposition, show large domains of the non-alloyed  $(p \times \sqrt{3})$ phase of the pure Bi/Ag(111) system coexisting with a mixed $(\sqrt{3} \times \sqrt{3})R30^{\circ}$ surface alloy (see Fig. S2 of Supplemental Material). These results indicate that when it is no longer possible to accommodate all the deposited atoms in substitutional positions of the top layer of Ag(111), the Bi atoms tend to separate from Sb to form the unalloyed Bi-pure $(p \times \sqrt{3})$ structure, while all Sb atoms remain in substitutional sites.

\begin{figure}
\includegraphics[width=1.0\columnwidth] {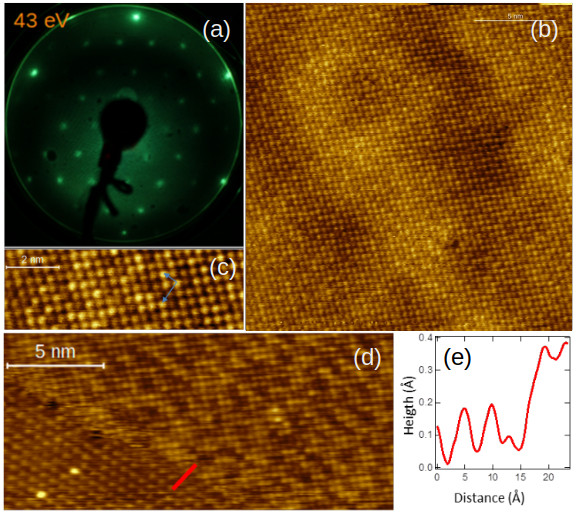}
 \caption{\label{BiSb_SbAg2} (a): LEED pattern taken at 43 eV corresponding  to 1/6ML of Bi deposited on a surface completely covered with the SbAg$_2$-Ag(111) surface alloy. (b), (c) and (d) STM images corresponding to the LEED pattern shown in (a). T. Cond.: (0.2V/1.4nA) for (a) and (b)  and  (0.5V/0.5nA) for  (d). (e): Line profile indicated as a red line in (d). }
 \end{figure}

On the other hand, if we begin with a Ag(111) surface fully covered by the SbAg$_2$ surface alloy and then deposit Bi, there is not enough space for all the Sb atoms to remain in the surface alloy, resulting in the formation of mixed ($3 \times \sqrt{3}$) rectangular structures.
Fig.~\ref{BiSb_SbAg2} illustrates the case for 1/6~ML of Bi deposition on the  SbAg$_{2}$, corresponding to a total amount of deposited atoms of 1/2~ML. Both the LEED pattern (Fig.~\ref{BiSb_SbAg2}(a)) and the atomic resolution STM images (Figs.~\ref{BiSb_SbAg2}(b) and (c)) indicate the formation of a commensurate ($3 \times \sqrt{3}$) rectangular structure with 4 atoms per unit cell. In addition to the rectangular phase, important areas of the surface are covered with the  $(\sqrt{3} \times \sqrt{3})R30^{\circ}$ surface alloy (see Fig.~S3 of Supplemental Material and related discussion). The STM image in Figure~\ref{BiSb_SbAg2}(d) shows also the interface between a $(\sqrt{3} \times \sqrt{3})R30^{\circ}$ surface alloy domain and a $(3 \times \sqrt{3})$ phase domain. The profile in Fig.~\ref{BiSb_SbAg2}(e) shows that the difference in height between these two domains is lower than $0.3$~\AA, indicating that both structures are supported on the same Ag(111) terrace.

The area covered by the rectangular structure grows with Bi deposition, until a value of 1/3~ML when this phase covers all the surface (see Fig. S4 of Supplemental Material and the related discussion). At this point, we have a total Sb+Bi deposition of 2/3~ML, and all the deposited atoms must be incorporated into the rectangular phase, with an average of two Sb atoms and two Bi atoms per unit cell. Moreover, although it is not straightforward to distinguish between Bi and Sb atoms from the STM image in Fig. \ref{BiSb_SbAg2}(c), we consistently observed a heterogeneous distribution of darker and brighter atoms. This suggests that the Sb and Bi atoms are distributed among the ($3 \times \sqrt{3}$) structure sites in a disorderly manner. In conclusion, the results indicate the formation of a rectangular non-alloyed phase, analogous to the one described above for the Sb/Ag(111) system. Remarkably, the same mixed structure can be obtained if the roles of Bi and Sb are inverted, i.e., starting from a surface covered with the BiAg$_2$ surface alloy and then depositing 1/3~ML of Sb (see the STM images in Figure S5 of the Supplemental Material). These conclusions are also supported by the XPS data, as we shall discuss  in the next paragraph.

Figure \ref{XPS3} compares the XPS spectra of Sb-3d$_{5/2}$ and Bi-4f$_{7/2}$ obtained from two surfaces with 1/3 ML of Sb and 1/3 ML of Bi, prepared using two alternative pathways. The spectra corresponding to the initial surface alloys are also included. A remarkable coincidence is observed between the spectra obtained from the two mixed ($3 \times \sqrt{3}$) surfaces, supporting the equivalence between the surfaces obtained through the two alternative pathways. Table \ref{table:XPS} summarizes the positions and areas resulting from fitting the spectra with single D-S lines. Further details on the fitting analysis can be found in Section S4 of the Supplemental Material.

Another important comparison involves the spectra obtained from the mixed ($3 \times \sqrt{3}$) surfaces and those measured on the surface alloys. First, the peak areas associated with the mixed ($3 \times \sqrt{3}$) surfaces remain practically unchanged with respect to those of the corresponding surface alloys, providing additional support for the proposed structural model where both Bi and Sb atoms are positioned in the top layer. Furthermore, both peaks related to the mixed ($3 \times \sqrt{3}$) structures shift towards higher binding energies compared to those of the surface alloys. Specifically, the Bi-4f${7/2}$ peak shifts by 0.14 eV, a value comparable to that associated with the dealloying transition in the case of the Bi-pure system \cite{Pussi2012}. Following the same trend, the Sb-3d${5/2}$ peak also exhibits a smaller shift towards higher binding energies. These observed binding energy shifts provide additional support for the hypothesis of a non-alloyed overlayer composed only of Bi and Sb atoms \cite{Pussi2012}.

%compares the XPS Bi-4f$_{7/2}$ spectrum for the BiAg$_2$ surface alloy (black) with the one for a total Sb+Bi deposition of 2/3~ML (green).
%We can observe that there is a displacement of the Bi-4f$_{7/2}$ peak position towards higher BEs of 0.2 eV, very similar to that observed in the case of the Bi-pure system in the de-alloying transition \textbf{ (INCLUIR REFERENCIA: K H L Zhang et al 2012 J. Phys.: Condens. Matter 24 435502 )}. There is also a small shift  in the Sb-3d$_{5/2}$ peak position, similar to the one observed during the dealloying transition in the pure Sb/Ag(111) system.
%These results point to the formation of a rectangular non-alloyed phase analogous to the one described above for the Sb/Ag(111) system.
%
%Moreover, although it is not straightforward to distinguish between Bi and Sb atoms from the STM image in Fig. \ref{BiSb_SbAg2}(c), we systematically observed a heterogeneous distribution of darker and brighter atoms, suggesting that the Sb and Bi atoms are distributed among the ($3 \times \sqrt{3}$) structure sites in disorderly manner.
%The same mixed structure can be obtained if the roles of Bi and Sb are inverted, i.e. starting from a surface covered with the BiAg$_2$ surface alloy and then depositing 1/3~ML of Sb (see the STM images in Figure S5 of Supplemental Material).

%The XPS spectra in red of Figures \ref{XPS}(a) and \ref{XPS}(b) correspond to this inverse preparation pathway.  The remarkable coincidence in both the Bi-4f$_{7/2}$ and Sb-3d$_{5/2}$ lines support the equivalence between the surfaces obtained following the two alternative pathways.

\begin{figure}
\includegraphics[width=1.0\columnwidth] {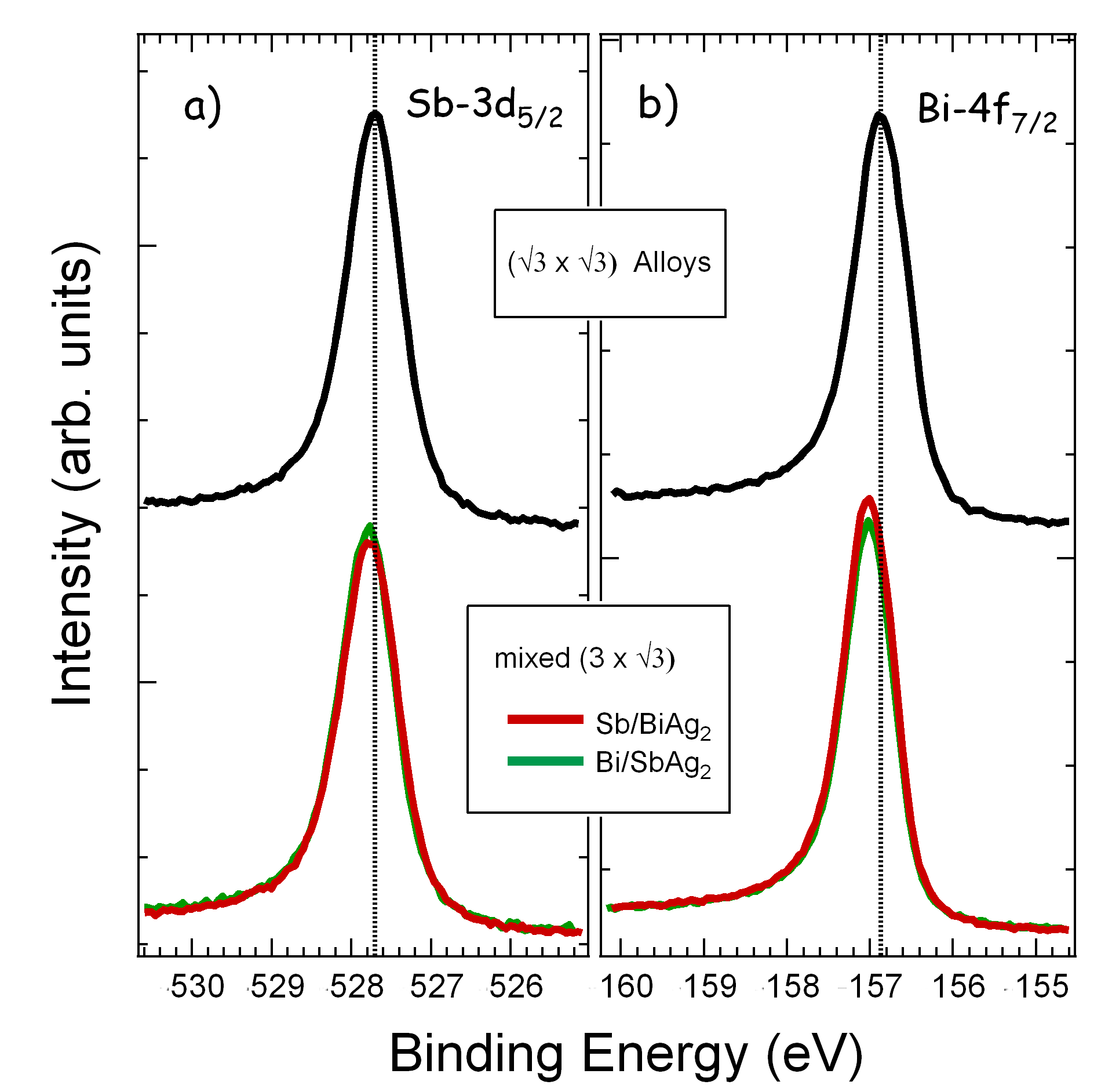}
 \caption{\label{XPS3}  (a) and (b) compare the XPS spectra derived from the Sb-3d$_{5/2}$  and Bi-4f$_{7/2}$  lines, respectively,  measured on two mixed ($3 \times \sqrt{3}$) surfaces with 1/3 ML of Sb and 1/3 ML of Bi, that were prepared  by following the two alternative preparation pathways.  The red spectra  correspond to depositing Sb on the BiAg$_2$ alloy whereas  the green ones correspond to the inverse preparation.  The top black spectra correspond to the initial surface alloys are also given (black). The corresponding BE positions and areas are summarized in Table \ref{table:XPS}. }
 \end{figure}

\begin{table}[h!]
\centering
\scalebox{0.7}{
\begin{tabular}{|l|c|c|c|c|}
\hline
 & \multicolumn{2} {c|} {Sb-3d$_{5/2}$}  &  \multicolumn{2}{c|} {Bi-4f$_{7/2}$} \\
Structure  & BE[eV]. & Area & BE[eV] & Area \\
\hline \hline
 BiAg$_2$ surface alloy &-& - & 156.83  & 1.00  \\
 SbAg$_2$ surface alloy &  527.69 & 1.00 & - & -  \\
mixed ($3 \times \sqrt{3}$), Sb/BiAg$_2$  (red) & 527.74 & 1.00 & 156.97  & 1.01  \\
mixed ($3 \times \sqrt{3}$), Bi/SbAg$_2$  (green) & 527.76  & 0.99  & 156.97 & 0.95 \\

\hline
\end{tabular}
}
\caption{BE Positions and areas of the D-S lines resulting from the fittings of the peaks shown in Fig. \ref{XPS3}. The areas of the Sb-4d$_{5/2}$ (Bi-4f$_{7/2}$) peaks measured from the mixed $(3 \times \sqrt{3})$ surfaces are normalized to that measured from the SbAg$_2$ (BiAg$_2$) surface alloy. Refer to Section S4 of the Supplemental Material for further details. }  \label{table:XPS}
\end{table}

In order to investigate the thermal stability of the mixed BiSb phase, we performed additional XPS, STM and UPS experiments. Figures \ref{XPS4}(a) and \ref{XPS4}(b) illustrate the evolution with the annealing temperature of the Sb-3d$_{5/2}$ and the Bi-4f$_{7/2}$  XPS peaks, respectively, for a surface completely covered with the $(3 \times \sqrt{3})$-Bi$_2$Sb$_2$ structure (obtained depositing 1/3~ML of Sb on the BiAg$_2$ surface alloy).  The heating of the sample was carried out gently using a radiative filament located under the sample. We found a notable general effect: temperature causes the reduction of the Sb peak but does not alter the Bi peak. In particular, when the temperature reached $(200 \pm 10)^{\circ}$C the area of the Sb-3d$_{5/2}$ peak was reduced to 55\% of the initial value, while the Bi-4f$_{7/2}$ spectra remained unchanged. In this manner, the total amount of atoms (Sb+Bi) has been reduced to values in the range between 1/3 and 2/3 ML, which opens the possibility, as discussed at the beginning of this section, for the formation of pure phases. In fact, this was confirmed on a sample annealed to $200^{\circ}$C, by means of STM and angle-resolved UPS experiments which indicate the appearance of the SbAg$_2$ surface alloy (see Fig. S8 of Supplemental Material). On the other hand, if we deposit an amount of Bi greater than 1/3~ML on a complete SbAg$_2$ surface alloy, the amount of Sb can be reduced up to a final Bi+Sb coverage equal to 2/3~ML, avoiding in this way the formation of the SbAg$_2$ surface alloy. Indeed, we performed XPS and UPS experiments which indicate that the SbAg$_2$ alloy is not formed after annealing (see the Figs. S9 and S10(b) of Supplemental Material for more details). We therefore conclude that temperature causes a selective effect on the atoms deposited on the surface, allowing the ratio of Sb to Bi to be varied in a controlled manner, and with the possibility to generate surfaces fully covered with Bi-rich mixed rectangular structures.

\begin{figure}
\includegraphics[width=1.0\columnwidth] {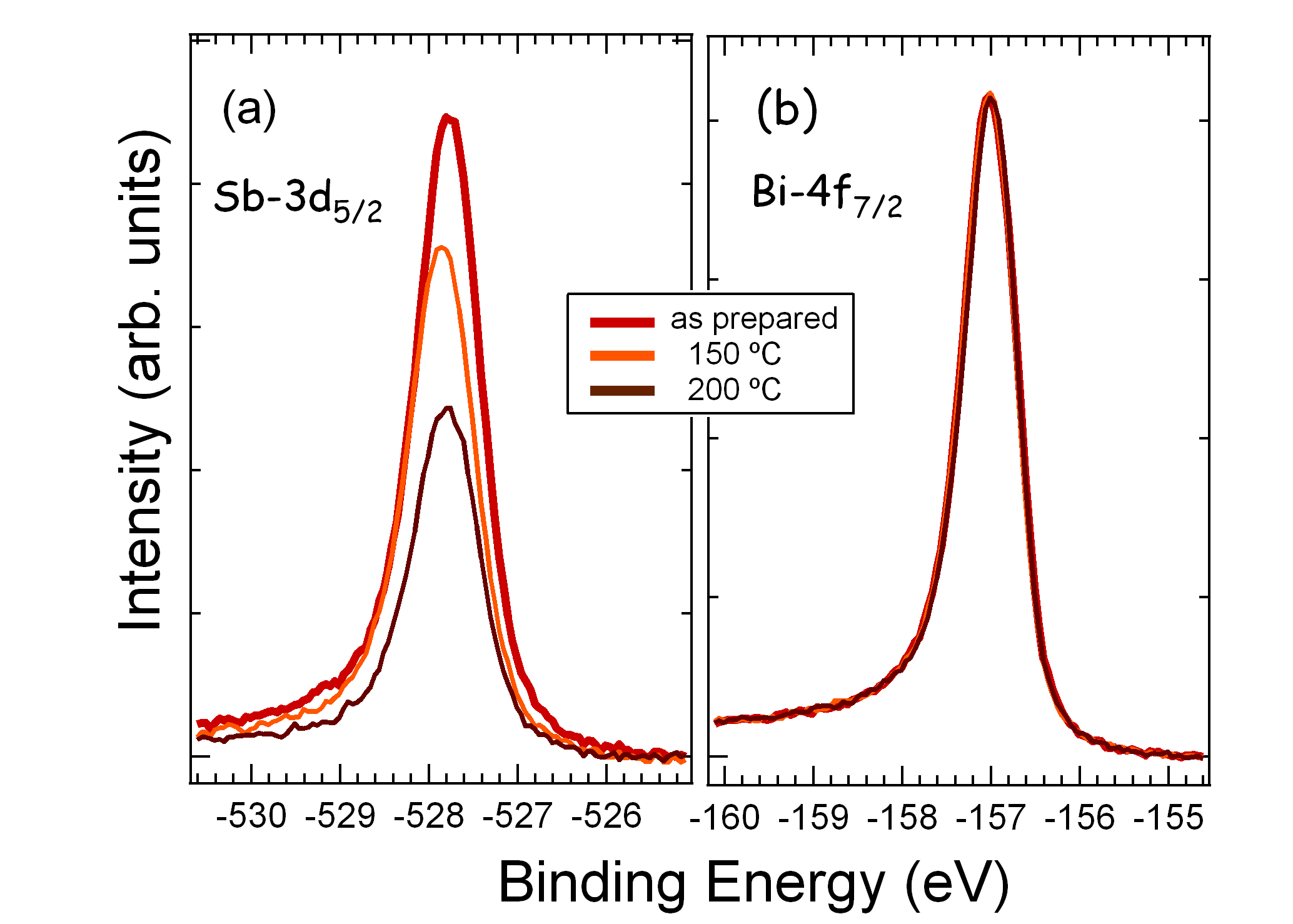}
 \caption{\label{XPS4}  (a) and (b) compare the XPS spectra derived from the Sb-3d$_{5/2}$  and Bi-4f$_{7/2}$ lines, respectively,  resulting from heating an independent sample prepared with 1/3 ML of Sb on a complete BiAg$_2$ surface alloy. The initial spectra are indicated with red while those obtained after heating are indicated in orange and brown.}
 \end{figure}

\begin{figure*}[t]
\includegraphics[width=0.9\textwidth]{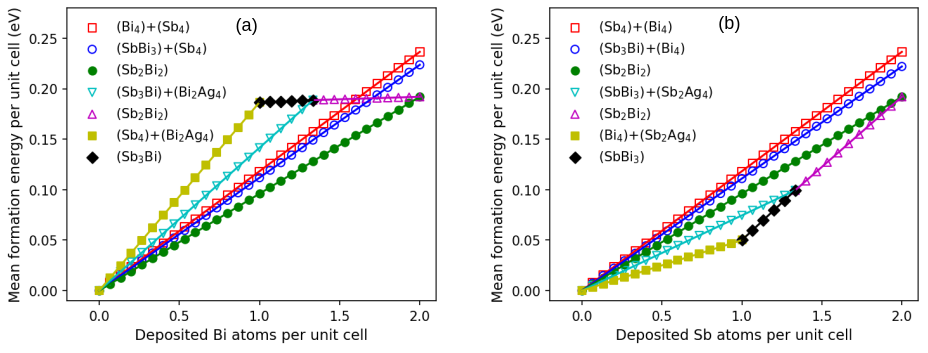}
 \caption{\label{fig:enef} Mean formation energy per unit cell as a function of amount of deposited Bi or Sb atoms, starting from: (a) SbAg$_2$ surface alloy, and (b) BiAg$_2$ surface alloy. Each symbol correspond to one of the reaction pathways (see section S8 of Supplemental Material for a description of the reactions), and the legend shows the products in the corresponding reaction. The formulas correspond to the possible structures within the $(3 \times \sqrt{3})$  unit cell, and we included the surface alloys as (Bi$_2$Ag$_4$) and (Sb$_2$Ag$_4)$.}
\end{figure*}

We performed total energy DFT calculations to improve our understanding of the relative stability of the possible structures that can appear during growth. We considered the SbAg$_2$ and BiAg$_2$ surface alloys, and several non-alloyed structures with 4 atoms within the ($3 \times \sqrt{3}$) unit cell: Sb$_4$, Sb$_3$Bi, Sb$_2$Bi$_2$, SbBi$_3$ and Bi$_4$. The last structure, Bi$_4$ on the ($3 \times \sqrt{3}$) Ag(111) unit cell, can be considered an approximation to the uncommensurate ($p \times \sqrt{3}$) phase observed for the pure Bi overlayer on Ag(111). See section S7 of Supplemental Material for details of the relaxed structures. In order to compare the stability of the different structures, we defined the relative formation energy $E_F$ for a given configuration as
\begin{equation}
  \label{eq:Ef}
  E_F = \left(E_{\rm conf}+ N_{\rm Ag} E_{\rm Ag}^{\rm (bulk)} \right) - \left(E_{\rm clean}^{(3\times\sqrt{3})} + N_{\rm Sb} E_{\rm Sb}^{\rm (alloy)} + N_{\rm Bi} E_{\rm Bi}^{\rm (alloy)}\right),
\end{equation}
where $E_{\rm conf}$ is the total energy of the corresponding configuration, $E_{\rm clean}^{(3\times\sqrt{3})}$ the total energy of a clean surface with the same unit cell, $N_{\rm Ag}$ is the number of substituted surface Ag atoms which are assumed to go to the bulk with the corresponding $E_{\rm Ag}^{\rm (bulk)}$ energy, $N_{\rm Sb}$ and $N_{\rm Bi}$ are the number of Sb and Bi atoms in the structure. As reference energy for these adsorbed atoms, we considered the corresponding surface alloy and defined, with X=Sb or Bi,
\begin{equation}
  \label{eq:EX}
  E_{\rm X} = \left(E_{\rm XAg_2}+ E_{\rm Ag}^{\rm (bulk)} \right) - E_{\rm clean}^{(\sqrt{3}\times\sqrt{3})},
\end{equation}
where $E_{\rm XAg_2}$ is the total energy per unit cell of the corresponding surface alloy and $E_{\rm clean}^{(\sqrt{3}\times\sqrt{3})}$ the total energy of a clean surface with the same unit cell.

The obtained relative formation energies per ($3 \times \sqrt{3}$) unit cell are: 0.373~eV for Sb$_4$, 0.283~eV for Sb$_3$Bi, 0.192~eV for Sb$_2$Bi$_2$, 0.149~eV for SbBi$_3$, and 0.100~eV for Bi$_4$. These values indicate that the richer in Bi a structure is, the lower its energy cost, which explains the experimental observations remarkably well. In fact, the high energy cost of the rectangular structure of pure Sb, i.e. the Sb$_4$ case, explains the difficulties encountered in achieving experimentally large areas of this phase. Furthermore, the high stability of the SbAg$_2$ surface alloy and the low formation energy of the Bi-pure rectangular structure will lead to a system composed of 1/6~ML of Sb and  1/3~ML of Bi, to separate into the Bi-pure rectangular phase and the SbAg$_2$  surface alloy, in line with our experimental observations.\footnote{The obtained STM images strongly suggest the formation of a mixed surface alloy in place of the predicted Sb-pure one. However, we interpret that this a metastable structure that forms due to kinetic limitations.} On the other hand, for the system with 1/3 ML of Sb and 1/3 ML of Bi the energetically more favorable is the mixed Sb$_2$Bi$_2$ structure while the formation energy of the other possible combinations of phases are within a range of $45$~meV per unit cell (see Supplemental Material for details). This explains the experimental observations that there is no pure phases separation and that there is a configuration disorder in the Sb and Bi atoms positions, as observed in the experimental STM images.

We then considered the relative formation energies as a function of coverage, for different possible Bi-Sb structures, when depositing one of the elements on a substrate fully covered with the surface alloy of the other element. Figure~\ref{fig:enef}(a) and (b) show the mean formation energy per ($3 \times \sqrt{3}$) unit cell when starting with a substrate fully covered with the SbAg$_2$ and the BiAg$_2$  surface alloy, respectively. See Tables~S2 and S3 in section S8 of the Supplemental Material for details on the considered reaction pathways. As we already mentioned above, when a total of 2/3~ML of Sb and Bi atoms has been deposited, corresponding to 4 atoms per ($3 \times \sqrt{3}$) unit cell, the more stable configuration is in both cases the Sb$_2$Bi$_2$ overlayer. However, the evolution of the system as a function of the amount of deposited atoms depends on the initial surface alloy. Indeed, when the substrate is initially covered with the  SbAg$_2$ surface alloy (Fig.~\ref{fig:enef}(a)), the most favorable reaction pathway (filled green circles) is formation of the Sb$_2$Bi$_2$ structure, independently of the amount of deposited Bi atoms. However, when starting with the  BiAg$_2$ surface alloy, the most favorable reaction pathway depends on the amount of deposited Sb atoms. Up to 1 Sb atom per unit cell (equivalent to 1/6~ML), the most favorable reaction pathway (yellow filled squares) is the formation of the pure Bi$_4$ structure with all the incoming Sb atoms going to the surface alloy.
When the amount of Sb reaches 1 atom per unit cell, there is no more place on the surface to continue the initial reaction. Therefore, the system begins the formation of the SbBi$_3$ at the expense of structures Bi$_4$ and SbAg$_2$ (indicated by black diamond). This reaction pathway continues up to 4/3 Sb atoms per unit cell, when two-thirds of the surface is covered by the SbBi$_3$ structure, while the remaining one-third is covered by the Sb/Ag(111) surface alloy. For higher Sb deposition, the only possibility is to begin the formation of the Sb$_2$Bi$_2$ structure at the expense of the SbBi$_3$ and Sb/Ag(111) phases.

Moreover, we observe that the curves in Fig.~\ref{fig:enef}(b) offer a theoretical description of the evolution of the (Sb$_2$Bi$_2$)/Ag(111) surface with temperature. Indeed, it was experimentally determined that the main  effect of temperature is to reduce the amount of Sb, which corresponds to run these curves from right to left. For instance, in the range of Sb coverage between 1/6 (1 atom per unit cell) and 1/3 ML (2 atoms per unit cell), it would not be a pure Sb$_2$Bi$_2$ structure but a mixture of this one with the Bi-rich SbBi$_3$, coexisting with the SbAg$_2$ surface alloy.

Finally, we considered the possibility of a large Rashba effect on the mixed Bi-Sb layer. The spin-orbit splitting of the electronic states on a surface depends proportionally on the SOC and the potential gradient $\nabla V$, with contributions both in- and out-of-plane.\cite{Premper2007}  An analysis of the atomic geometries of the calculated mixed structures (Sb$_3$Bi, Sb$_2$Bi$_2$, and SbBi$_3$) shows that the larger Bi atoms prefer to be at a higher $z$-position with respect to the substrate (see Fig.~S12 and related text in Section S7 of the Supplemental Material). The resulting layer corrugation is larger than 0.25~\AA\ for the three mixed structures. Therefore, the combination in this system of a high SOC for Bi atoms, a large in-plane $\nabla V$ due to lighter Sb neighbors, and a significant out-of-plane $\nabla V$ due to layer corrugation, allows us to expect a more important Rashba effect than in other similar systems, such as a Bi monolayer on Cu(111).\cite{Mathias2010}

\section{Conclusions}

We have shown that single layers of a mixed Bi$_{(1-x)}$Sb$_x$ composition can be grown on the Ag(111) surface. The preparation procedure is  simple: it is  indispensable to start from a surface completely covered by either of the two  pure surface alloys, the SbAg$_2$ or the BiAg$_2$ one, and to deposit the other element on it. The obtained atomic geometry of the mixed structure is very similar to the one we observed for Sb/Ag(111): an unalloyed overlayer with rectangular ($3 \times \sqrt{3}$) structure and four atoms inside the unit cell, forming zig-zag chains. When the total (Bi+Sb) coverage is 2/3~ML, the mixed Bi-Sb phase completely covers the surface. Moreover, we determined that in the Bi$_{(1-x)}$Sb$_x$ layer, the Bi and Sb atoms are distributed disorderly in the structure's sites.

The energetics derived from the DFT calculations indicates that the SbAg$_2$ surface alloy has a high stability, and that as the Bi content increases in an overlayer structure, its energy cost decreases. Indeed, these calculations predict a pathway to the formation of Bi-rich non-alloyed phases: starting with the $(\sqrt{3} \times \sqrt{3})R30^{\circ}$ Bi/Ag(111) surface alloy and depositing less than 2/9~ML of Sb.

The  mixed ($3 \times \sqrt{3}$)-Sb$_2$Bi$_2$ phase remains stable up to approximately $(150 \pm 10)^\circ$C, at which point the diffusion of the Sb atoms to the Ag bulk becomes activated, resulting in a reduction of the surface Sb. Starting with a surface completely covered with the Sb$_2$Bi$_2$ phase, the reduction of Sb leads to the formation of Bi-rich ($3 \times \sqrt{3}$) mixed structures and simultaneously triggers the reappearance of the SbAg$_2$ surface alloy. Therefore, a pathway to obtain a surface completely covered with a Bi-rich ($3 \times \sqrt{3}$) mixed structure, is to deposit an amount of Bi greater than 1/3~ML onto a substrate completely covered with SbAg$_2$ surface alloy and subsequently reduce the amount of Sb to achieve a final Bi+Sb coverage of 2/3~ML.

The obtained mixed Bi-Sb structures have all the necessary ingredients, high SOC for Bi atoms and large in- and out-of-plane potential gradient, to expect that this system could present an important Rashba effect.

\section{Acknowledgments}

We acknowledge the financial support from CONICET (Grant PIP-2021-1404)

\end{document}